\documentclass[runningheads]{llncs}
\usepackage[T1]{fontenc}
\usepackage{lmodern}
\usepackage{amsmath}
\usepackage{booktabs}
\usepackage{graphicx}

\PassOptionsToPackage{hyphens}{url}
\usepackage{url}

\usepackage{algorithmic}
\usepackage{multicol}
\usepackage{array}
\usepackage{placeins}
\usepackage{tcolorbox}
\usepackage{tikz}
\usepackage{listings}
\usepackage{hyphenat}
\usepackage{tabularx}
\usepackage{xspace}
\usepackage[shortlabels]{enumitem}
\usepackage{pifont}
\usepackage{colortbl}
\usepackage{multirow}
\usepackage{longtable}
\usepackage[utf8]{inputenc}
\usepackage{xcolor}
\usepackage{microtype}
\usepackage{subcaption}
\usepackage{hyperref}
\usepackage{macros}
\DeclareRobustCommand{\perm}[1]{\texttt{\footnotesize #1}}

\makeatletter
\g@addto@macro{\UrlBreaks}{\UrlOrds}

\def\@citex[#1]#2{\if@filesw\immediate\write\@auxout{\string\citation{#2}}\fi
	\@tempcnta\z@\@tempcntb\m@ne\def\@citea{}\@cite{\@for\@citeb:=#2\do
		{\@ifundefined
			{b@\@citeb}{\@citeo\@tempcntb\m@ne\@citea\def\@citea{, }{\bfseries
					?}\@warning
				{Citation `\@citeb' on page \thepage \space undefined}}%
			{\setbox\z@\hbox{\global\@tempcntc0\csname b@\@citeb\endcsname\relax}%
				\ifnum\@tempcntc=\z@ \@citeo\@tempcntb\m@ne
				\@citea\def\@citea{, }\hbox{\csname b@\@citeb\endcsname}%
				\else
				\advance\@tempcntb\@ne
				\ifnum\@tempcntb=\@tempcntc
				\else\advance\@tempcntb\m@ne\@citeo
				\@tempcnta\@tempcntc\@tempcntb\@tempcntc\fi\fi}}\@citeo}{#1}}
\def\@citeo{\ifnum\@tempcnta>\@tempcntb\else
	\@citea\def\@citea{, }%
	\ifnum\@tempcnta=\@tempcntb\the\@tempcnta\else
	{\advance\@tempcnta\@ne\ifnum\@tempcnta=\@tempcntb \else
		\def\@citea{--}\fi
		\advance\@tempcnta\m@ne\the\@tempcnta\@citea\the\@tempcntb}\fi\fi}
\makeatother

\title{Silent Consent, Persistent Risk: Android Permission Groups and Custom Permissions}
\titlerunning{Silent Consent, Persistent Risk}
\author{Olawale Amos Akanji\orcidID{0009-0005-7419-5108}$^{\ding{41}}$ \and Manuel Egele \and Gianluca Stringhini}
\authorrunning{O. A. Akanji et al.}
\institute{Boston University, Boston MA 02215, USA\\
	\email{\{olawalea, megele, gian\}@bu.edu}}

\begin{document}

	\maketitle              
	\begin{abstract}

		Android's permission system is designed to balance usability with informed consent, yet two legacy mechanisms still undermine that balance in Android~16: (i) permission groups that silently auto-grant new permissions within a group after a user's initial approval, and (ii) \texttt{normal}-level custom permissions that are auto-granted at install and enable cross-app access with no user visibility. We conduct a longitudinal analysis of 19.3~million APKs spanning 5.97~million unique apps (distinct package identifiers) from the AndroZoo repository, combined with on-device validation on Android~16. Among 2{,}244{,}575 multi-version apps, 381{,}026 (17\%) silently gain permissions within already-granted groups. Using VirusTotal detections with primary threshold $t=20$, apps flagged as malware expand within groups at a higher rate than benign apps (odds ratio $=1.35$, $p<0.001$); the association holds across every tested threshold and concentrates in permission-heavy apps ($\text{OR}=2.06$ in the top quartile). We also identify 307 cross-developer \perm{normal}-custom-permission pairs that expose contacts, SMS, location, authentication credentials, user identity, and medical records to unrelated apps without any user prompt. A lightweight prototype built on public Android APIs recorded 23 silent expansion events across 13 apps during a 96-day single-device pilot, showing that update-time transparency is reachable without OS modification. Our results show that consent erosion persists despite a decade of platform hardening and affects apps ranging from obscure utilities to widely deployed and pre-installed software.

		\keywords{Android Security \and Mobile Privacy \and Cross-App Data Exposure \and Permission Systems \and Permission Groups \and Custom Permissions \and Informed Consent}
	\end{abstract}

	\section{Introduction}
	\label{sec:introduction}
	Android powers over three billion active devices~\cite{android3billion}, and its permission system governs how apps access sensitive device resources and user data, including contacts, location, camera, microphone, and storage. Apps declare the permissions they require in their manifest, and the system enforces access control at runtime. Permissions fall into three protection levels: \perm{normal} permissions (e.g., internet access) are auto-granted at install because they pose minimal risk; \perm{dangerous} permissions (e.g., reading contacts) require explicit user consent because they access private data; and \perm{signature} permissions are granted only to apps signed with the same developer key as the app that defined them. This model gives users visibility into what resources apps can access and, for dangerous permissions, the ability to grant or deny access.
	
	Modern permission systems must balance two competing goals: protecting users from privacy violations while enabling seamless application interaction. Android embodies this tension in its permission architecture. Early versions (Android~1.0--5.1) relied on install-time permission lists that users largely ignored, leading to ``symbolic consent''~\cite{androidPermOverview,Felt,felt2,kelley2012conundrum}. Android~6.0 addressed this by introducing runtime permissions: apps must now request each dangerous permission individually at the moment the protected resource is needed, and users see a contextual dialog they can grant or deny.
	
	Despite this shift, two mechanisms inherited from Android~1.0 remain active in Android~16 and continue to weaken the consent guarantees that runtime prompts were designed to provide: \textbf{permission groups} and \textbf{normal-level custom permissions}.
	
	\textbf{Permission Groups.} Android clusters related permissions into named groups to reduce the number of runtime prompts shown to users~\cite{androidpermissiongroups}. Once a user grants any permission from a group, the system silently auto-grants (i.e., automatically approves without showing a prompt) subsequent additions to that same group during app updates, enabling \emph{silent capability expansion} without renewed consent.
	
	\textbf{Normal-Level Custom Permissions.} Android allows developers to define proprietary permissions that guard exported components~\cite{Gamba2024Mules,liAndroidCustomPermissions2021}. When declared with protection level \texttt{normal} (the default), any requesting app, regardless of developer identity, receives automatic approval at install time with no user prompt and no indication in Settings, creating an \emph{implicit permission delegation} path that bypasses Android's dangerous-permission enforcement. Section~\ref{sec:motivation} examines the concrete security tensions and attack scenarios these mechanisms enable.

	Prior work has documented confused-deputy attacks and custom-permission misuse enabling cross-app leakage~\cite{Autogranted,Gamba2024Mules,liAndroidCustomPermissions2021,Tuncay}: Calciati et al.~\cite{Autogranted} measured permission-group auto-grant prevalence on 2{,}865{,}553 app versions, and Gamba et al.~\cite{Gamba2024Mules} analysed \perm{normal}-level custom-permission exposure over 2.2M apps. Both mechanisms persist in Android~16 substantially unchanged. We revisit them at ecosystem scale, validate their exploitability on Android~16, and evaluate a lightweight public-API transparency intervention.

	We investigate four research questions:
	
	\begin{itemize}[leftmargin=*,noitemsep,topsep=3pt]
		\item \textbf{RQ1.} To what extent do permission groups enable silent expansion of app capabilities across updates without renewed user oversight?
		\item \textbf{RQ2.} Is silent permission-group expansion an ecosystem-wide consent failure, or is it disproportionately an attribute of malware rather than benign apps?
		\item \textbf{RQ3.} How prevalent are \perm{normal}-level custom permissions that enable cross-developer data access, what sensitive data do they expose, and can we confirm active exploitability on a current Android release?
		\item \textbf{RQ4.} Can lightweight, interface-level interventions restore visibility into update-time permission changes without compromising usability?
	\end{itemize}

	To address these questions, we re-estimate both mechanisms on Android~16 at ecosystem scale (19{,}324{,}760 APKs from 5{,}965{,}583 unique apps in AndroZoo, deduplicated by package name), compare expansion rates between VirusTotal-flagged and benign apps under a principled sensitivity sweep, validate \perm{normal}-custom-permission re-delegation on-device through per-pair Probe Apps, attribute the recovered exploitation surface to app-core code versus bundled third-party libraries, and release a runnable update-time transparency prototype built on public Android APIs. Our contributions are:
	
	\begin{enumerate}[leftmargin=*,noitemsep,topsep=3pt]
		\item We show that silent permission-group expansion remains a systemic consent failure on Android~16: 17\% of multi-version apps in the AndroZoo corpus silently gain permissions within already-granted groups, and on-device testing confirms that the auto-grant mechanism still fires across every dangerous group on Android~16 with no runtime dialog or Settings indication.
		
		\item We show that this silent expansion is not a uniformly latent design weakness but is preferentially engaged by malware: the odds-ratio association is statistically significant at every tested VirusTotal threshold in the Zhu et al.~\cite{zhu2020measuring} range and, after stratifying by permission breadth, concentrates in permission-heavy apps ($\text{OR}=2.06$ in the top quartile).
		
		\item We identify 307 cross-developer \perm{normal}-custom-permission pairs that re-delegate contacts, SMS, location, authentication credentials, user identity, and medical records, and confirm active exploitability on Android~16 by instrumenting Probe Apps that retrieve the sensitive payloads; call-site attribution assigns the majority of the recovered surface to bundled third-party libraries rather than to app-core code.
		
		\item We implement a public-API update-time transparency prototype and validate it in a 96-day Pixel~7 deployment: 23 silent expansion events across 13 apps, including apps with over one billion installs, at roughly one notification every four days, showing that lightweight interface-level interventions can restore update-time consent visibility without OS modification.
	\end{enumerate}

	\section{Motivation}
	\label{sec:motivation}
	Despite Android's shift to runtime permissions and richer controls, two legacy mechanisms, permission groups and \perm{normal}-level custom permissions, continue to undermine informed consent. Permission groups allow silent capability expansion within already-approved groups during app updates, while \perm{normal} custom permissions enable cross-app access to guarded components without any user prompt or runtime dialog.
	
	\subsection{Google's Justification: Prompt Fatigue and Openness}
	Android's permission grouping model minimizes repeated dialogs by auto-granting future permissions within a group once any member is approved, framing this as a response to prompt fatigue~\cite{androidpermissiongroups}.
	
	For custom permissions, the rationale is interoperability. Developers define proprietary permissions to guard exported components~\cite{androidCustomPerms}, each carrying a protection level: \perm{normal} (auto-granted at install, the default), \perm{signature} (same-key apps only), or \perm{dangerous} (runtime user consent). The \perm{signature} level restricts grants to apps signed with the same key, though key sharing across organizations weakens this guarantee~\cite{Gamba2024Mules}. By contrast, \perm{normal}, the default when developers omit the level, provides no protection: any requesting app receives automatic approval regardless of developer identity, with no prompt and no Settings entry. Gamba et al.~\cite{Gamba2024Mules} documented real apps exposing private data via \perm{normal} custom permissions in a responsible disclosure; Google acknowledged the issue but framed it as a consequence of Android's openness~\cite{Gamba2024Mules}, leaving a recognized yet unresolved attack surface.

	\subsection{The Security Tension in Modern Android}
	Permission groups and custom permissions each introduce distinct security tensions that persist on modern Android. We examine three facets: risk asymmetry from forced bundling within groups, granularity that the platform offers at grant time but hides at revocation, and the trust boundary that normal-level custom permissions silently bypass.
	
	\noindent\textbf{Risk Asymmetry and Forced Bundling.} Despite recent granularity improvements (Android~13 split media access into separate \perm{READ\_MEDIA\_IMAGES}, \perm{READ\_MEDIA\_VIDEO}, and \perm{READ\_MEDIA\_AUDIO}~\cite{androidgranular}; Android~14 added a partial photo-picker~\cite{android14PartialAccess}), many permission groups still bundle capabilities with fundamentally different risk profiles. For example, a two-factor authentication app that requests \perm{READ\_SMS} to extract one-time passcodes may later add \perm{SEND\_SMS} in an update. Because both belong to the SMS group, Android auto-grants the addition without any prompt. If the app is compromised or acquired by a malicious actor, it can silently dispatch premium-rate messages, a qualitative expansion from passive read access to active send capability that operates entirely within documented group semantics.
	
	\begin{figure}[htbp]
		\centering
		\begin{subfigure}[b]{0.45\columnwidth}
			\centering
			\includegraphics[width=\linewidth,height=5cm,keepaspectratio]{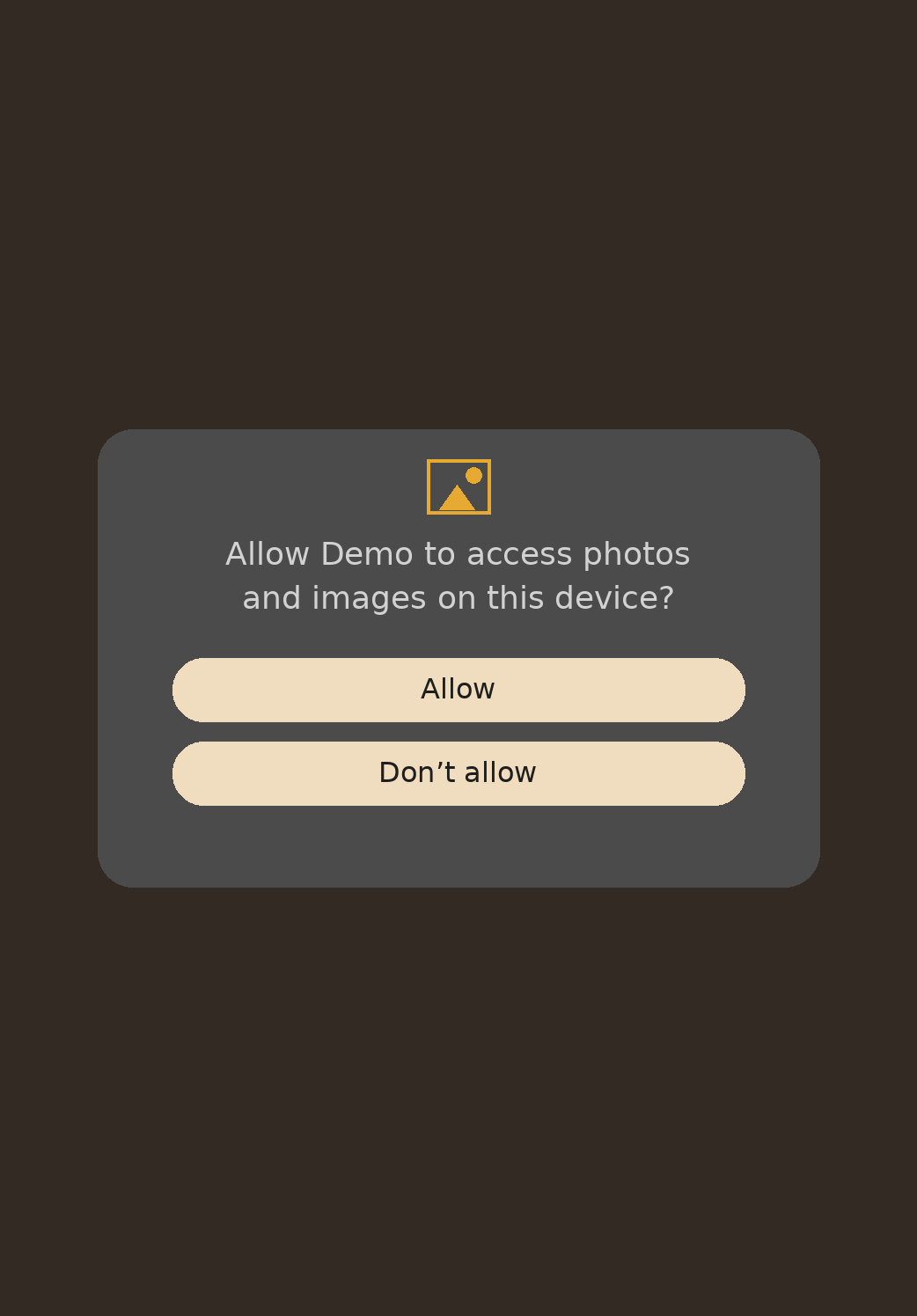}
			\caption{Grant-time prompt}
			\label{fig:images}
		\end{subfigure}
		\hfill
		\begin{subfigure}[b]{0.45\columnwidth}
			\centering
			\includegraphics[width=\linewidth,height=5cm,keepaspectratio]{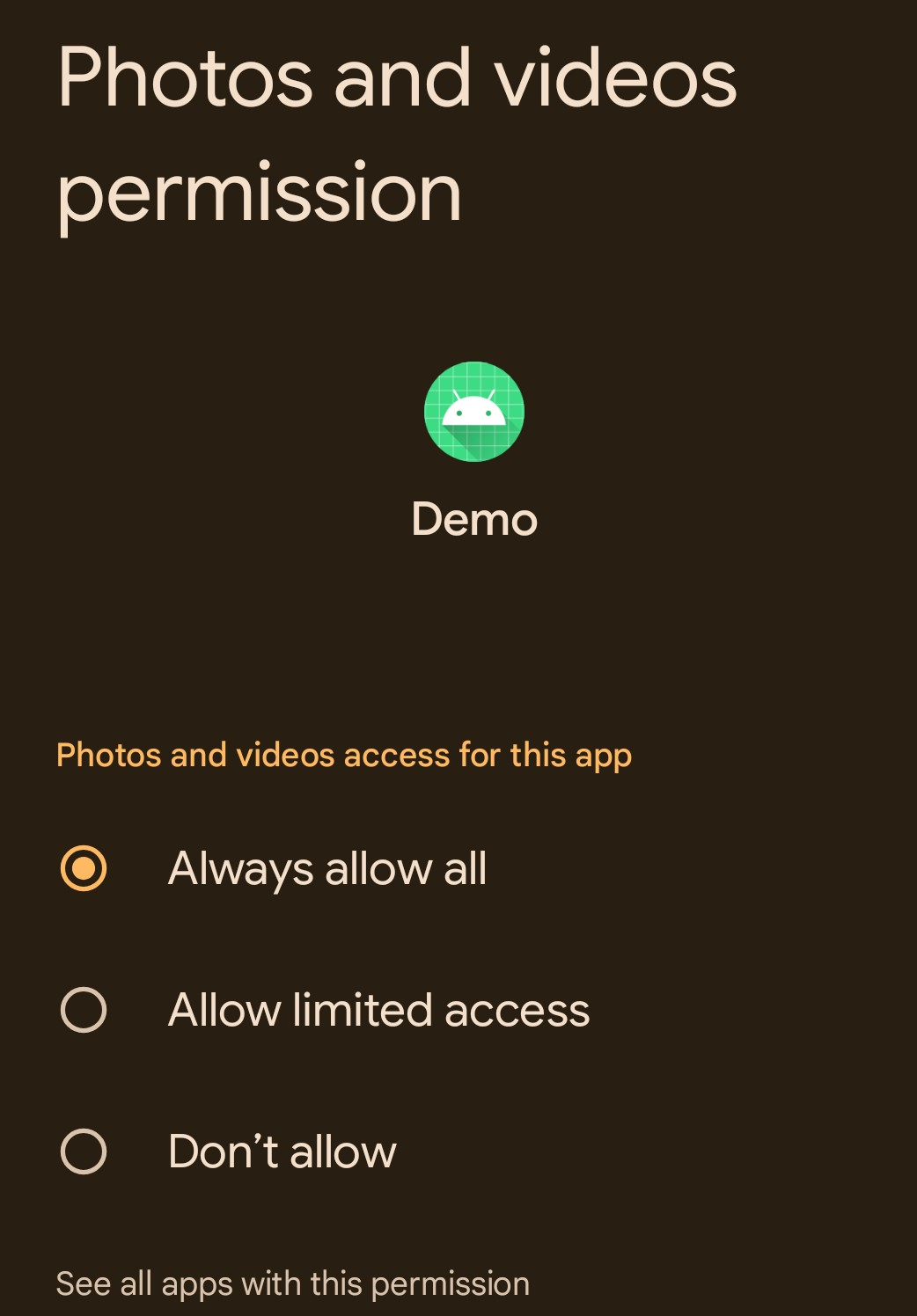}
			\caption{Settings toggle}
			\label{fig:combined}
		\end{subfigure}
		\caption{Android~16 media permissions: (a)~the runtime dialog offers a granular prompt for \texttt{READ\_MEDIA\_IMAGES}, but (b)~Settings collapses all media permissions into a single ``Photos and videos'' toggle, erasing prompt-level granularity. \texttt{READ\_MEDIA\_IMAGES} is a standalone permission that excludes video and audio; a user who granted only photo access has no way to revoke video access independently because Settings does not expose the per-permission granularity that the runtime prompt created.}
		\label{fig:media-permissions}
	\end{figure}

	\noindent\textbf{Granularity Without Revocation.} Android has introduced fine-grained prompts but not equally fine-grained revocation, creating a one-way ratchet: permissions can be silently widened through updates, but users cannot individually narrow them. Media permissions illustrate the gap: an app requesting only \perm{READ\_MEDIA\_IMAGES} triggers a single photo-access prompt~\cite{androidphotoandvideo}, but if a later update adds \perm{READ\_MEDIA\_VIDEO}, Android auto-grants it silently. As Figures~\ref{fig:images} and~\ref{fig:combined} show, Settings provides only a single ``Photos and videos'' toggle. Revoking video necessarily revokes photos, forcing users to choose between over-permissioning and losing functionality entirely~\cite{nielsen-heuristics,Wijesekera}.

	\noindent\textbf{The Limits of ``Openness'' as Defense.}
	Permission groups at least provide a one-time consent decision; the problem is that silent updates can expand capabilities without renewed approval. \perm{normal}-level custom permissions provide no user-facing consent at all. A developer who defines a custom permission must declare four manifest attributes (\texttt{android:name}, \texttt{android:protectionLevel}, \texttt{android:label}, and \texttt{android:description}), explicitly signaling intent to restrict access. Yet the default protection level (\perm{normal}) nullifies that intent: every requesting app receives automatic approval at install time, with no prompt and no Settings entry. We refer to the app that defines a \perm{normal}-guarded exported component as the \emph{exploitable app}, and the unrelated third-party app that declares the same permission name to reach that component as the \emph{exploiting app}.
	
	Our threat model is an exploiting app from a different developer that declares the exploitable app's \perm{normal} custom permission and invokes exported components guarded by it. For example, a health-tracking app storing medical records via a \perm{normal}-guarded provider can have its data silently read by any other app that declares the same permission name, without the exploitable developer's knowledge, turning interoperability into an unmediated delegation channel.

	\section{Methodology}
	\label{sec:method}
	We combine longitudinal static analysis of 19.3M APKs (5.97M unique apps, deduplicated by package name) spanning 2013--2025 with temporal analysis of permission-group expansion trends aligned to major Android policy changes, malware vs benign comparison using VirusTotal classification, cross-developer custom permission linking through bytecode inspection of sensitive data flows, dynamic validation on Android~16 devices, and a 96-day deployment of a transparency prototype built using only public APIs.
	
	\subsection{Dataset}
	\label{subsec:dataset}
	We analyse 19{,}324{,}760 APKs from 5{,}965{,}583 unique packages in AndroZoo~\cite{allix2016androzoo} spanning Google Play and alternative markets, extracting requested permissions, custom permission definitions, protection levels, and exported components from each {\footnotesize\texttt{AndroidManifest.xml}}.
	
	\subsection{Permission Groups: Longitudinal Analysis and Dynamic Testing}
	\label{subsec:group_method}
	We perform a fully automated longitudinal analysis over the corpus using Androguard~\cite{androguard} to parse {\footnotesize\texttt{AndroidManifest.xml}} from every APK and extract its \perm{<uses-permission>} declarations. For each package with at least two versions, we compare adjacent versions, map every declared permission to its Android Open Source Project (AOSP) protection group, and flag cases where the later version introduces a permission whose group was already present in the earlier version; under Android's runtime model, once any member of a group has been granted, additional members from that group are implicitly approved.
	
	We also aggregate expansion activity by release year to determine whether major platform changes reduced prevalence or merely shifted which groups are affected (Figure~\ref{fig:temporal_timeline}).
	
	To assess whether silent permission-group expansion is disproportionately associated with malware, we label each app using VirusTotal (VT) detection counts, that is, the number of antivirus engines that classify a given APK as malicious when it is submitted to the VirusTotal aggregator. Zhu et al.~\cite{zhu2020measuring} show that threshold-based aggregation is meaningful over the range $2 \leq t \leq 39$, with thresholds below~15 offering the strongest precision--recall trade-off. We use $t=20$ as the primary threshold and report sensitivity at $t \in \{2,5,10,39\}$ to span the admissible range.
	
	For each app we record whether it ever exhibited a permission-group expansion, producing a 2$\times$2 table of expanding vs.\ non-expanding apps crossed with malware vs.\ benign labels. From this table we compute the odds ratio~(OR)~\cite{agresti2007introduction}, where values $>1$ indicate positive association, a Pearson chi-squared test for significance~\cite{pearson1900chi}, and a Mantel--Haenszel stratified estimate~\cite{mantel1959statistical} that controls for the total number of permissions each app declares.
	
	Finally, we verify that these silent expansions actually occur on a current device. On physical Android~16 devices we install minimal test apps: a baseline version requests a single group permission (e.g., {\footnotesize\texttt{READ\_CONTACTS}}); a follow-up version adds another from the same group (e.g., {\footnotesize\texttt{WRITE\_CONTACTS}}). After granting the baseline, we update in place and observe whether Android surfaces a new runtime dialog or silently enables the added capability. We also inspect Settings to determine whether revocation is exposed per permission or only via a group-level toggle. This pairing of longitudinal analysis with device-level observation lets us measure prevalence while establishing behavioral ground truth on Android~16.

	\subsection{Custom Permissions: Detection, Linking, and Validation}
	\label{subsec:custom_method}
	We implement the custom-permission pipeline as a fully automated corpus-scale analysis,\footnote{Artifact availability: the analysis scripts, Probe App harness, AutoGrant Monitor APK artifact, and aggregate per-group outputs are available at \url{https://github.com/marshalwahlexyz1/Silent-Consent-Persistent-Risk-Android-Permission-Groups-and-Custom-Permissions.git}.} using Androguard~\cite{androguard} for manifest and DEX parsing. Each APK passes through three static stages: manifest classification and component-to-permission binding, bytecode-level data-sensitivity analysis of the declaring side, and call-site resolution on the requesting side. These stages produce candidate pairs, which we then filter to those signed by different developer certificates and validate end-to-end on an Android~16 device.
	
	\emph{Manifest classification.} We classify every \perm{<permission>} element against the canonical Android Open Source Project (AOSP) permission list parsed from successive Android releases; declarations outside that list are labeled custom, and we record the \perm{android:protectionLevel} (defaulting to \perm{normal} when omitted). We then resolve each custom permission to the exported component it guards (\perm{<provider>}, \perm{<activity>}, \perm{<service>}, or \perm{<receiver>}) and restrict to attachments that are simultaneously \perm{protectionLevel="normal"} and \perm{exported="true"}, since same-process guards and \perm{signature}-protected components cannot be re-delegated across developer boundaries.
	
	\emph{Bytecode-level data-sensitivity analysis.} Manifest matching alone yields many benign candidates whose guarded components expose only app configuration or user-interface state. We therefore analyse the DEX bytecode of each declaring component for evidence that it returns AOSP-protected data, following prior DEX-level Android inspection work~\cite{akanji2026cost,Enck}. For every declaring component we use Androguard's~\cite{androguard} method- and basic-block-level analysis layer to walk return paths: for content providers we resolve \texttt{getType()} and trace \texttt{ContentProvider.query()} return values back to sensitive sources such as \perm{ContactsContract} and \perm{Telephony.Sms}, and extract the column-name string constants they expose (e.g., \texttt{PHONE\_NUMBER}, \texttt{USER\_EMAIL}); for activities and receivers we trace payloads set via \texttt{setResult()}; for services we resolve method signatures returned by \texttt{onBind()}~\cite{Gamba2024Mules}.
	
	\emph{Call-site resolution on the requesting side.} A bare \perm{<uses-permission>} declaration is insufficient evidence of access: prior measurement work on Android has consistently had to recover the actual invocation sites before drawing behavioural conclusions, whether for permission re-delegation~\cite{Gamba2024Mules,Tuncay}, covert data access~\cite{reardon50ways}, or taint tracking of sensitive sources~\cite{Enck}. For every app that declares a custom permission matching a sensitive provider, we use Androguard's~\cite{androguard} DEX analysis layer to enumerate each method's invocation list and flag calls to \texttt{ContentResolver.query}/\texttt{insert}/\texttt{update}/\texttt{delete}/\texttt{call}; we then resolve the authority-URI argument via constant propagation through \texttt{Uri.parse}, \texttt{String} concatenation, and \texttt{StringBuilder} chains, and accept the call site only when the resolved authority matches the target provider. For each matched call we record the enclosing method's fully qualified declaring class so that downstream app-core vs.\ third-party library attribution can be performed directly on the bytecode-observable call graph. Because custom-permission names, provider authorities, and \perm{android:permission} manifest attributes are runtime-load-bearing strings and must survive identifier renaming to function at all, manifest-level pair identification is obfuscation-stable; call-site enumeration is the only stage where ProGuard repackaging reduces recall, and since every reported pair is anchored on an observed call site, obfuscation cannot inflate precision. It can only cause us to miss pairs whose exploiting call site has been renamed beyond recognition, making the reported set a lower bound.
	
	\emph{Runtime validation.} Because static analysis establishes only that a cross-developer data path exists, we validate each pair whose provider exposes named columns in bytecode via an on-device experiment on Android~16. For each such pair we instrument both halves. On the exploitable side we install the declaring app on the device and grant its AOSP dangerous permission (e.g., \perm{READ\_CONTACTS}); the app queries the corresponding system provider and stores the returned rows inside its own exported \texttt{ContentProvider}, which remains guarded only by the \perm{normal} custom permission. On the requesting side we build a minimal \emph{Probe App} that declares that custom permission via \perm{<uses-permission>} (auto-granted at install with no user prompt) and issues a \texttt{ContentResolver.query} against the exploitable app's provider authority using the statically resolved projection. A pair is confirmed only when (i)~the Probe App reads back a non-empty \texttt{Cursor} with valid field values from the exploitable app's provider, and (ii)~the Android \texttt{AppOps} journal (the platform's per-operation access log) records a protected operation such as \perm{OP\_READ\_CONTACTS} attributed to the exploitable app, confirming that the sensitive data behind the \perm{normal} guard originated from an AOSP dangerous-permission read rather than from test fixtures.
	
	\subsection{Prototype: Surfacing Silent Expansions to Users}
	\label{subsec:prototype}
	
	We implement a lightweight prototype that detects silent permission expansions without OS modification. The prototype registers for package lifecycle broadcasts (\perm{ACTION\_PACKAGE\_ADDED}, \perm{ACTION\_PACKAGE\_REPLACED}), extracts each app’s declared permissions at update time, and compares them against a cached snapshot from the previous version. It then maps newly introduced permissions to their groups and triggers a notification only when a new permission appears within an already granted group. To reduce noise, we deduplicate per app/version, ignore first-time group introductions, and include a direct link to the app’s permission settings. This provides update-time visibility with minimal user interruption while relying only on public APIs.
	\section{Results}
	\label{sec:results}
	We first present the prevalence, temporal evolution, and malware association of permission-group expansion. We then turn to cross-developer exploitation through \perm{normal} custom permissions and report the deployment results of the update-time transparency prototype.
	\begin{table}[t]
		\centering
		\caption{Top-6 permission-group expansion flows (main groups).}
		\label{tab:topk_flows_main}
		\scriptsize
		\begin{minipage}[t]{0.48\linewidth}
			\centering
			\setlength{\tabcolsep}{3pt}
			\begin{tabular}{l l r}
				\toprule
				Group & Flow & Count \\
				\midrule
				CONTACTS & Get Accts $\rightarrow$ Read & 13,164 \\
				CONTACTS & Read $\rightarrow$ Get Accts & 5,830 \\
				CONTACTS & Read $\rightarrow$ Write & 4,747 \\
				CONTACTS & Get Accts $\rightarrow$ Write & 4,299 \\
				CONTACTS & Write $\rightarrow$ Get Accts & 1,999 \\
				CONTACTS & Write $\rightarrow$ Read & 552 \\
				SMS & Send $\rightarrow$ Receive & 2,186 \\
				SMS & Send $\rightarrow$ Read & 2,152 \\
				SMS & Receive $\rightarrow$ Read & 1,302 \\
				SMS & Receive $\rightarrow$ Send & 1,236 \\
				SMS & Read $\rightarrow$ Send & 1,057 \\
				SMS & Read $\rightarrow$ Receive & 1,007 \\
				\bottomrule
			\end{tabular}
		\end{minipage}\hfill
		\begin{minipage}[t]{0.48\linewidth}
			\centering
			\setlength{\tabcolsep}{3pt}
			\begin{tabular}{l l r}
				\toprule
				Group & Flow & Count \\
				\midrule
				PHONE & Read State $\rightarrow$ Call & 26,430 \\
				PHONE & Call $\rightarrow$ Read State & 9,674 \\
				PHONE & Read State $\rightarrow$ Read Num. & 6,115 \\
				PHONE & Read State $\rightarrow$ Process Out. & 4,605 \\
				PHONE & Call $\rightarrow$ Read Num. & 2,928 \\
				PHONE & Call $\rightarrow$ Process Out. & 2,306 \\
				CALL LOG & Read $\rightarrow$ Write & 206 \\
				CALL LOG & Write $\rightarrow$ Read & 43 \\
				\bottomrule
			\end{tabular}
		\end{minipage}
	\end{table}
	\subsection{Permission Groups: Longitudinal Prevalence and Dynamic Confirmation}
	\subsubsection{Longitudinal Analysis.}
	Of 2{,}244{,}575 multi-version apps, 381{,}026 (17.0\%) silently gain at least one permission inside an already-granted group, contributing 928{,}452 auto-granted additions in total, or 2.44 additions per expanding app on average. Table~\ref{tab:topk_flows_main} lists the dominant observed flows for the CONTACTS, SMS, PHONE, and CALL\_LOG groups, while Appendix Table~\ref{tab:topk_flows_appendix} reports the corresponding top flows for the remaining five groups. This rate is of the same order as the earlier Calciati et al.~\cite{Autogranted} measurement,\footnote{Cross-study rate comparisons should be treated as approximate because datasets, years, and denominators differ.} indicating that the mechanism has persisted on modern Android.
	
	Per-group expansion volume (Appendix~\ref{app:per_group}, Table~\ref{tab:per_group}) shows that STORAGE, LOCATION, and PHONE account for 94.5\% of all expansion cases. The communication groups, SMS and CALL\_LOG, contribute a much smaller share of total expansions but, as we show below, have the highest concentration of malware-flagged expanders.
	
	We next test whether the aggregate rate could be inflated by AndroZoo's multi-market collection model. Using the AndroZoo \texttt{markets} metadata, we stratify expanding apps by market provenance and separate same-market histories from cross-market histories. The effect remains visible in the Play-only and non-Play-only buckets and is not concentrated in cross-market package histories, which indicates that the result is not driven primarily by market-specific version mixing.
	
	\subsubsection{Temporal Trends and Platform Evolution.}
\begin{figure*}[t]
		\centering
		\includegraphics[width=\textwidth]{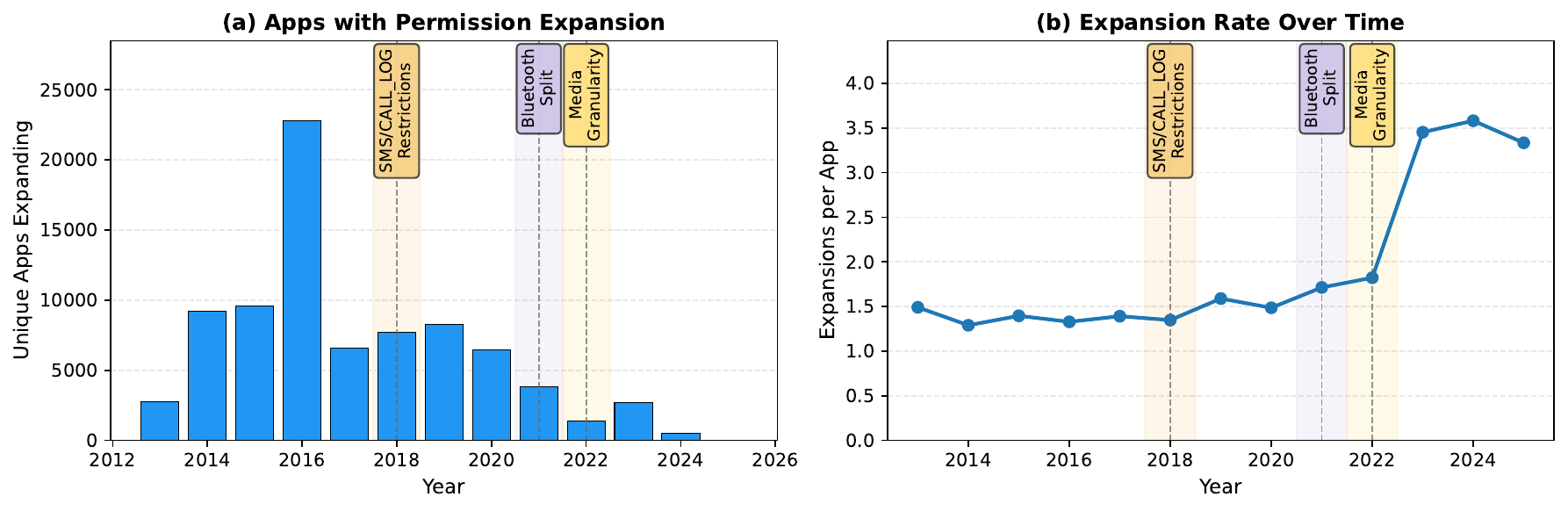}
		\caption{Permission group expansion trends over time (2013--2025) with major Android platform changes annotated. (a) Number of unique apps exhibiting at least one permission-group expansion per year. (b) Expansion rate (average expansions per expanding app) per year.}
		\label{fig:temporal_timeline}
	\end{figure*}

	Examining ecosystem evolution shows that expansion behavior varies across periods. Based on each app's release date, Figure~\ref{fig:temporal_timeline} summarizes yearly trends. Panel~(a) shows that the number of unique apps exhibiting expansion peaked around 2016 ({\raise.17ex\hbox{$\scriptstyle\sim$}}22,000 apps) and has declined sharply in recent years ({\raise.17ex\hbox{$\scriptstyle\sim$}}500--2,700 apps from 2022--2024). Panel~(b) shows that the expansion rate (average expansions per expanding app) remained relatively stable (1.3--1.9) through 2021 but increased sharply to 3.5+ in 2023--2024. This divergence indicates that while fewer apps now expand permissions, those that do are adding more permissions per update. We interpret these trends in light of two kinds of platform change: changes that alter group composition directly, and adjacent privacy changes that leave within-group auto-grant semantics intact.
	
	\noindent\emph{Group-affecting changes.}
	\begin{itemize}[leftmargin=*,noitemsep,topsep=3pt]
		\item \textbf{CALL\_LOG separation from PHONE (2018).} Android 9 split \perm{READ\_CALL\_LOG}, \perm{WRITE\_CALL\_LOG}, and \perm{PROCESS\_OUTGOING\_CALLS} into a new CALL\_LOG group~\cite{android9CallLogGroup}; Google Play further restricted SMS/Call Log access to default handler apps~\cite{googlePlaySmsCallLogPolicy}. SMS/CALL\_LOG expansions drop sharply by 2020 (155 events, --80.0\% vs 2017).
		\item \textbf{NEARBY\_DEVICES group (2021--2022).} Android 12 introduced a \perm{NEARBY\_DEVICES} group for Bluetooth~\cite{android12BluetoothPermissions}; Android 13 added \perm{NEARBY\_WIFI\_DEVICES}~\cite{android13BehaviorChanges}. Expansions remain low ($\leq$44/year).
		\item \textbf{Media granularity (2022).} For API 33+, Android 13 replaces \perm{READ\_EXTERNAL\_STORAGE} with the granular media permissions \perm{READ\_MEDIA\_IMAGES}, \perm{READ\_MEDIA\_VIDEO}, and \perm{READ\_MEDIA\_AUDIO}~\cite{androidmedia,android13BehaviorChanges}. The legacy permission nevertheless remains in STORAGE, creating a reverse expansion path: an app granted only \perm{READ\_MEDIA\_IMAGES} can silently gain \perm{READ\_EXTERNAL\_STORAGE} in a later update. STORAGE expansions dip in 2022 (1{,}073 vs 2{,}985, --64.1\%) but spike in 2023 (8{,}159).
		\item \textbf{Background body sensors (2022).} Android 13 added \perm{BODY\_SENSORS\_BACKGROUND}~\cite{android13BehaviorChanges}. SENSORS expansions remain rare (0.3\% of events).
	\end{itemize}
	
	\noindent\emph{Adjacent changes.} Runtime permissions (Android~6.0)~\cite{android6RuntimePerms}, one-time grants and auto-reset (Android~11)~\cite{android11PrivacyPermissions,androidAutoResetBlog}, and system pickers (SAF~\cite{androidSAF}, Contact Picker~\cite{androidContactPicker}, Photo Picker~\cite{androidPhotoPicker,androidStorageExperienceBlog}) strengthen consent in complementary ways but do not alter within-group auto-grant semantics. Users still receive no notification when an app adds a permission to an already-approved group. The temporal analysis therefore points to reactive, per-group policy responses that redraw the boundaries of silent expansion without addressing the underlying mechanism.
	
	\subsubsection{Dynamic Confirmation on Android~16.}
	We verify these patterns on Android~16 (API~36) through controlled experiments. For each of the nine dangerous permission groups, we install paired test applications: the baseline version requests a single group member (e.g., \perm{READ\_MEDIA\_IMAGES}); the update adds an additional permission from the same group (e.g., \perm{READ\_MEDIA\_VIDEO}). After granting the baseline permission and updating in place, Android~16 automatically grants every newly introduced group permission in all nine groups: no runtime dialog, no update notification, and no distinction in Settings. These experiments confirm that the longitudinal expansion patterns represent real, silent capability growth on the current platform. These results answer \textbf{RQ1}: permission groups enable widespread silent expansion---17\% of multi-version apps gain permissions within already-granted groups---and the auto-grant mechanism remains active across all nine dangerous groups on Android~16.
	
	\subsection{Malware vs Benign Apps}
	We compare how often malware and benign apps use permission-group expansion to determine whether the mechanism is primarily a malware tactic or a broader ecosystem problem. Using VirusTotal labels (Section~\ref{sec:method}), we classify each app by its highest detection score across all versions, with $t = 20$ as the primary threshold.
	
\begin{figure}[t]
		\centering
		\includegraphics[width=0.85\columnwidth]{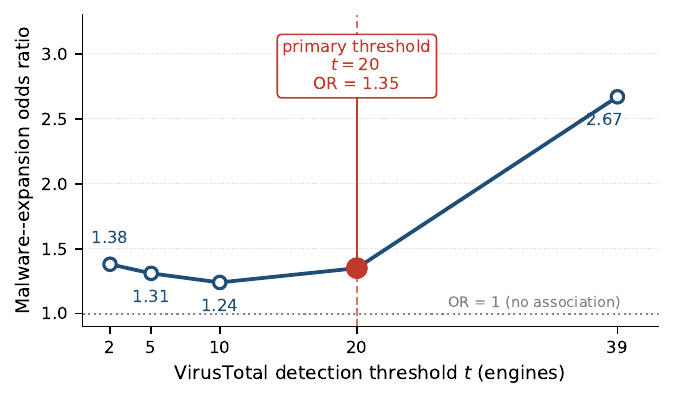}
		\caption{Malware--expansion odds ratio across the five reported VirusTotal thresholds (primary threshold $t=20$ highlighted).}
		\label{fig:vt_sensitivity}
	\end{figure}

	\noindent\textbf{Per-app, malware-flagged apps expand at a higher rate than benign apps.} We call an app \emph{malware-flagged} at threshold $t$ when at least $t$ of the VirusTotal antivirus engines classify any version of that app as malicious. At $t = 20$, 1.63\% of the 381{,}026 expanding apps are malware-flagged, compared with 1.29\% of the 2{,}244{,}575 multi-version baseline. The resulting odds ratio (OR) is 1.35 ($\chi^2 = 430.9$, $p < 0.001$), so malware-flagged apps are more likely than benign apps to have silently expanded within a permission group. Figure~\ref{fig:vt_sensitivity} shows that this is not a one-threshold artifact: the association remains statistically significant at every tested value, with OR $=$ 1.38 at $t{=}2$, 1.31 at $t{=}5$, 1.24 at $t{=}10$, 1.35 at $t{=}20$, and 2.67 at $t{=}39$ (all $p < 0.001$). In absolute terms, however, most expanding apps are still benign (98.37\% at $t = 20$) because benign apps dominate the ecosystem numerically.
	
	One possible explanation for this gap is that malware simply requests more permissions overall and therefore has more opportunities to expand. We test that possibility by stratifying apps by total declared permission count into four quartiles and recomputing the association (Table~\ref{tab:stratified_or}). The stratified OR is 1.05 (95\% CI [1.02, 1.08], $p = 0.001$), which is statistically significant but substantially attenuated. Permission breadth therefore explains most of the crude association. The remaining effect is concentrated in the top quartile ($\geq$24 permissions), where malware expands at about twice the rate of benign apps (OR $=$ 2.06), while the three lower quartiles show comparable rates.
	
	Not all permission groups carry equal malware risk. Decomposing the 381{,}026 expanding apps by group and overlaying the VT labels, the communication groups concentrate the highest malware share (SMS: 72.6\%, CALL\_LOG: 54.3\%), while the high-volume groups STORAGE and LOCATION contribute the largest absolute malware counts despite lower within-group percentages. To characterise what ``malware-flagged'' actually covers in our corpus, we retrieve per-engine family labels for all 6{,}225 apps that cross the $t=20$ threshold: the flagged bucket is a mixture of adware (36.1\%), trojans/malware (36.0\%), PUA/riskware (22.2\%), and other categories (5.6\%), so the odds-ratio association we report holds against a broad malware population rather than being driven by a single family. These results answer \textbf{RQ2}: silent expansion is an ecosystem-wide consent failure, but malware-flagged apps engage it at a higher per-app rate, with the association concentrated in permission-heavy apps.
	\subsection{Custom Permissions: Evidence of Real-World Exploitation}
	
	We investigate three aspects of \perm{normal}-level custom permissions on Android~16: (1)~prevalence and component types guarded, (2)~whether those components handle sensitive user data, and (3)~whether apps from unrelated developers exploit these permissions to access that data without user awareness.
	
	\subsubsection{Prevalence of Normal Custom Permissions.}
	Across the corpus we extract 2{,}630{,}134 distinct custom permission names (Table~\ref{tab:cusperm_levels}). Most use \perm{signature} protection (90.4\%), while 9.4\% are defined as \perm{normal} and are therefore auto-granted to any requesting app. Table~\ref{tab:cusperm_levels} also breaks the \perm{normal} subset down by the type of exported component each permission guards. Of these, only the 18{,}475 permissions attached to content providers expose persistent structured user data in our pipeline. In our sample, exported activities, services, and broadcast receivers return user-interface state, remote procedure call (RPC) primitives, or event broadcasts and, under our bytecode filter for AOSP-protected sources, yield no cross-developer pairs in which the exploitable app demonstrably returns sensitive fields. We therefore focus the remainder of this section on \perm{normal}-level \perm{<provider>} guards.
	
	\begin{table}[h]
		\centering
		\caption{Custom permissions: protection-level split (left) and, for the \perm{normal} subset, guarded component type (right).}
		\label{tab:cusperm_levels}
		\scriptsize
		\setlength{\tabcolsep}{4pt}
		\begin{minipage}[t]{0.48\linewidth}
			\centering
			\begin{tabular}{lrr}
				\toprule
				Protection level & Count & \% \\
				\midrule
				\perm{signature} & 2{,}378{,}863 & 90.4 \\
				\perm{normal}    & 246{,}958     & 9.4 \\
				\perm{dangerous} & 3{,}853       & 0.1 \\
				Other            & 461           & $<$0.1 \\
				\midrule
				\textbf{Total}   & \textbf{2{,}630{,}134} & \textbf{100.0} \\
				\bottomrule
			\end{tabular}
		\end{minipage}\hfill
		\begin{minipage}[t]{0.48\linewidth}
			\centering
			\begin{tabular}{lrr}
				\toprule
				Component (\perm{normal}) & Count & \% \\
				\midrule
				\perm{<provider>} & 18{,}475  & 7.5 \\
				\perm{<activity>} & 87{,}527  & 35.4 \\
				\perm{<service>}  & 4{,}005   & 1.6 \\
				\perm{<receiver>} & 4{,}743   & 1.9 \\
				Unattached        & 132{,}208 & 53.5 \\
				\midrule
				\textbf{Total}    & \textbf{246{,}958} & \textbf{100.0} \\
				\bottomrule
			\end{tabular}
		\end{minipage}
	\end{table}
	
	\subsubsection{Cross-Developer Pairs.}
	Among the 18{,}475 \perm{normal}-guarded providers, we focus on those that constitute an \emph{active} cross-developer exploitation surface, that is, providers for which an unrelated app both declares the matching permission and actually calls into it. We therefore filter this set to pairs in which (i)~an
	\emph{exploitable app} declares the \perm{normal} custom permission and
	exposes the guarded \texttt{ContentProvider}, (ii)~an \emph{exploiting app}
	declares the same permission name and issues a \texttt{ContentResolver}
	\texttt{query}/\texttt{insert}/\texttt{update}/\texttt{delete}/\texttt{call}
	against that authority, and (iii)~the two APKs are signed with different
	developer certificates. This filter yields 307 active cross-developer pairs over
	323 unique packages: 83 appear only as exploitable apps, 79 only as
	exploiting apps, and 161 in both roles across different pairs. These apps
	expose data through one \perm{normal} permission while accessing another
	app's data through a different one, so the exploitation surface is
	bidirectional.
	
	We characterize the 323 unique packages along three axes: certificate provenance, Google Play installation tiers, and, for the requesting side, whether the exploiting \texttt{ContentResolver} call originates from app-core code or from a bundled library.
	
	Grouping the 323 APKs by the public key of their \texttt{v2+} APK signing block, 228 are signed by third-party developer keys, 49 by original equipment manufacturer (OEM) platform keys (Samsung, OPPO, Motorola, Amazon, Microsoft), and 46 by Google platform keys; 74 of the 307 statically confirmed active cross-developer pairs have at least one OEM- or platform-signed APK on either side. Cross-referencing the same 323 packages against live Google Play metadata, 102 are currently listed on Play and 221 are not (removed, region-restricted, or only shipped through OEM or alternative channels). Within the 102 Play-listed packages, 24 are in the $\geq$100M-install tier: 12 appear only as exploitable apps, 4 only as exploiting apps, and 8 in both roles across different pairs. These results show that the cross-developer exploitation surface is not confined to obscure third-party apps; it includes OEM- and platform-signed packages and Play-distributed apps with large installation counts on both the defining and requesting sides.
	
	For each exploiting \texttt{ContentResolver} call site, we attribute the declaring class either to the exploiting app's own code or to a bundled third-party library, following prior Android work that recovers embedded libraries and uses them to study third-party code behavior in the wild~\cite{backes2016reliable,book2013case,grace2012unsafe,kollnigFaitAccompliEmpirical,li2017libd,ma2016libradar}. We classify a call site as \emph{app-core} when the declaring class shares a package prefix with the exploiting app's manifest package, as \emph{third-party library} when it matches a known SDK package prefix, and as \emph{unclassified} otherwise. ProGuard repackaging primarily moves third-party code into the unclassified bucket rather than into app-core, so the resulting counts are a conservative lower bound on SDK involvement. To cross-validate the package-prefix heuristic, we ran LibScout~\cite{backes2016reliable} profile matching against 8{,}542 pre-computed library signatures on the 329 pair APKs: 284 of 315 successfully analysed APKs (90.2\%) contain at least one signature-level library match (mean 24.0 libraries per APK), confirming that bundled third-party SDK code is pervasive in the pair corpus. Of the 273 pairs with a recoverable exploiting call site, 126 (46\%) resolve exclusively to third-party library code, 127 (47\%) contain both app-core and third-party call sites, 10 (4\%) resolve exclusively to app-core, and the remainder are unclassified.
	
	For 40 of the 307 statically confirmed active cross-developer pairs, bytecode analysis of the exploitable provider recovers readable column-name string constants, so only these pairs admit both field-level naming and a reproducible on-device probe; 39 of the 40 are still installable on Android~16, and we exclude the remaining pair.
	
	For each of the 39 dynamically validated pairs on Android~16, we compile a \emph{Probe App} that reproduces the exploiting call site while holding no dangerous permissions and execute it against the exploitable provider. All 39 probes retrieve the sensitive payload via \texttt{ContentResolver.query}, and the Android \texttt{AppOps} journal records a protected operation, confirming that the \perm{normal} custom permission functions as a runtime re-delegation channel on Android~16 for the subset amenable to field-level probing. On this dynamically validated subset, the same attribution yields 18 SDK-only, 19 mixed, 1 app-core-only, and 2 unclassified.
	
	For the remaining 267 statically confirmed pairs, we confirm the presence of a persistent structured store (201 SQLite-backed, 62 file-backed) but cannot recover field names or run an equivalent on-device probe. We therefore report 307 as the statically confirmed active cross-developer surface and restrict field-level and dynamic claims to the 39 pairs validated on Android~16. Table~\ref{tab:exploitation_cases} reports the exposed data categories for that 39-pair subset, with each pair assigned to its primary category so the pairs are counted exactly once.
	
	\subsubsection{Categories of Real-World Exploitation.}
	Whether these flows are intentional interoperability or accidental exposure is not a defense: in both cases, user data crosses developer boundaries without a runtime prompt, Settings visibility, or explicit user authorization. Table~\ref{tab:exploitation_cases} reports, for the 39 probed pairs, the primary data category returned by the exploitable provider and whether that category is also gated by an AOSP dangerous permission. Type~A categories duplicate AOSP-gated data through a \perm{normal}-only channel (so a \perm{normal}-permission holder reads data that the AOSP gate would otherwise prompt for); Type~B categories are app-specific records that no AOSP permission covers, so the \perm{normal} permission is the \emph{only} access control.
	
	\begin{table}[h]
		\centering
		\scriptsize
		\setlength{\tabcolsep}{4pt}
		\caption{Primary data category assigned to each of the 39 probed pairs. Type~A: also gated by an AOSP dangerous permission. Type~B: no AOSP gate; the \perm{normal} permission is the sole access control.}
		\label{tab:exploitation_cases}
		\resizebox{\linewidth}{!}{%
			\begin{tabular}{llrl}
				\toprule
				Type & Data Category & Pairs & AOSP Gate (Type~A only) \\
				\midrule
				A & Contact data                    & 12 & \perm{READ\_CONTACTS} \\
				A & Authentication credentials      & 8  & \perm{GET\_ACCOUNTS} \\
				A & User identity                   & 7  & \perm{READ\_PHONE\_STATE} \\
				A & Location                        & 2  & \perm{ACCESS\_FINE\_LOCATION} \\
				A & Messages                        & 1  & \perm{READ\_SMS} \\
				B & File and storage paths          & 5  & --- \\
				B & Medical / health records        & 1  & --- \\
				B & Financial data                  & 2  & --- \\
				B & Settings / config               & 1  & --- \\
				\midrule
				\textbf{Total category instances}   &     & \textbf{39} & \\
				\bottomrule
		\end{tabular}}
	\end{table}
	
	These results answer \textbf{RQ3}: \perm{normal}-level custom permissions are widespread (246{,}958 definitions, 9.4\% of all custom permissions), and among the 18{,}475 that guard content providers, 307 statically confirmed active cross-developer pairs expose contacts, SMS, location, credentials, identity, and medical records to unrelated apps. Of these, 39 dynamically validated pairs on Android~16 are confirmed exploitable on-device, and the majority of recoverable exploitation call sites originate from bundled third-party SDKs rather than app-core code.
	
	\subsection{Prototype Deployment: Feasibility of Update-Time Transparency}
	
	The prototype (Section~\ref{sec:method}) runs on a Google Pixel~7 with Android~16 and 214 installed applications spanning social media, finance, and utilities. Over a 96-day deployment it detected 23 silent expansion events across 13 apps, none of which surfaced any indication to the user that a new permission had been silently granted. The affected apps include widely-used third-party apps in social media (1B+ installs), short-video (1B+ installs), photo sharing (1B+ installs), and finance (10M+ installs) categories, as well as OEM/platform software responsible for connectivity, device diagnostics, and accessibility services. We report these in aggregate rather than by package name because disclosure remains active. Each expansion event triggers a notification such as \emph{``This app has gained a new permission: \textbf{Read Images}. Review in Settings.''} Without that notification, the new permission is silently active, with no consent dialog, no visual cue, and no Settings entry. The notification restores informed decision-making by letting the user revoke the group, keep the permission, or uninstall the app, though as discussed in Section~\ref{sec:motivation} Android exposes only a group-level toggle, so revoking one unwanted permission means losing the entire group.
	
	Notification burden remains low: 23 events over 96 days amounts to roughly one notification every four days. Our longitudinal data show a median rate of 1.0~within-group additions per app per year; even if every installed app expanded once per year, a user with 80~apps~\cite{buildfire2024apps} would receive about 1.5~notifications per week, while a power user with 365~apps would receive roughly one per day. This is well below the dozens of daily social-media alerts most users already manage, and each notification concerns a genuine capability change. These results answer \textbf{RQ4} as a feasibility existence proof: on a single device and install profile, lightweight, interface-level interventions can restore update-time transparency without imposing a per-day user burden; ecosystem-scale deployment and user-perception studies remain future work.
	
	\section{Discussion}
	\label{sec:discussion}
	
	Our measurements show that Android's response to both mechanisms has been selective rather than systematic: the platform has repeatedly added user-facing controls when individual risks became visible (call log, Bluetooth, broad media access), yet update-time group expansion and \perm{normal}-level custom-permission re-delegation still bypass consent at ecosystem scale on Android~16. For permission groups, the implication is straightforward: within-group additions should trigger risk-sensitive re-consent or, at minimum, explicit update-time disclosure and revocation controls that preserve the granularity offered at grant time. Our 96-day prototype shows that such disclosure is implementable on top of public APIs with a manageable notification budget.
	
	For \perm{normal} custom permissions, Google already documents the cross-app exposure risk, but governance has not translated into enforcement. Build-time lint, store-review gates on \perm{normal}-guarded sensitive exports, and a \perm{signature} default for developer-defined permissions would reduce this surface without breaking legitimate interoperability. The fact that most recovered exploitation call sites originate from bundled third-party SDKs rather than first-party app code also suggests that remediation should not stop at app developers; it should extend to SDK vendors and to store policies that flag SDK-introduced \perm{normal}-guarded exports during review. Questions that remain open, including how users interpret update-time notices at scale and when re-consent should be mandatory, require user studies and platform-policy experiments.
	
	\subsection{Limitations}
	\label{subsec:limitations}
	The permission-group analysis is subject to one dataset-level qualification: AndroZoo aggregates APKs from multiple markets, and for 22.77\% of the expanding apps the two consecutive versions we compared came from different markets. In those cases, an apparent expansion may reflect market-specific or regional variants rather than a user-observed in-place update; we therefore treat this subset explicitly as qualified and show separately that the effect persists in the Play-only and non-Play-only buckets. For the malware-association analysis, VirusTotal detection counts are a behavioral proxy rather than a ground-truth malware-family label, so our result is an association with aggregated VT flagging, not with any single malware family. For the custom-permission analysis, the 307-pair set is a static lower bound: reflection, dynamic loading, and native-code accesses are out of scope, and while manifest-level pair identification is obfuscation-stable, identifier renaming reduces requester-side call-site recovery. Exact field-level categorization is therefore available only for the 40 pairs whose exploitable-side \texttt{ContentProvider} exposes readable column-name strings, and our app-core versus library attribution is conservative because ProGuard repackaging can shift third-party code into the unclassified bucket. Finally, the 96-day Pixel~7 deployment of the transparency prototype is a feasibility study on a single device and install profile, not an ecosystem-scale estimate of notification burden.

	\section{Related Work}
	\label{sec:related}
	
	\noindent\textbf{Usable Security and Consent Mechanisms.}
	Early Android versions relied on install-time permission lists that users largely ignored. Felt et al.~\cite{Felt} showed that users rarely understood or acted on bulk permission requests, while Kelley et al.~\cite{kelley2012conundrum} found that even simplified permission displays failed to improve comprehension. These findings motivated Android's transition to runtime dialogs in Android 6.0. However, subsequent work revealed persistent challenges: Felt et al.~\cite{felt2} demonstrated that users struggled to make informed decisions even with contextual prompts, and Wijesekera et al.~\cite{Wijesekera} showed that user preferences for permission grants varied significantly based on context, with many grants later regretted. Research on permission decision-making has emphasized the gap between user mental models and actual system behavior. Almuhimedi et al.~\cite{almuhimediYourLocationHas2015} found that users were unaware of how frequently apps accessed location data, leading to surprise when shown access logs. Wijesekera et al.~\cite{wijesekeradynamicallygrantedpermissions} proposed dynamically adjusting permission grants based on user context, finding that static grants poorly matched user preferences. These studies frame permissions as \emph{interfaces for decision-making} where incomplete cues and prompt fatigue shape behavior, providing context for our investigation of silent post-grant expansion.
	
	\noindent\textbf{Permission Evolution and Capability Drift.}
	Longitudinal analyses have documented how Android's permission architecture produces ``capability drift,'' gaps between what users believe they approved and what apps can subsequently do. Wei et al.~\cite{wei2012permission} tracked permission evolution across Android versions, showing that apps accumulate permissions over time. Calciati et al.~\cite{Autogranted} demonstrated that once any group member is granted, later additions are automatically approved, eroding users' ability to track capability changes. Backes et al.~\cite{backes2016demystifying} analyzed the Android framework's permission enforcement, revealing inconsistencies between documented and actual behavior. Almomani and Al-Khayer~\cite{AlmomaniAndroidpermissionanalysis} argued that grouping by functionality rather than risk leads users to conflate distinct harms, undermining risk-aware consent. Reardon et al.~\cite{reardon50ways} documented how apps circumvent permission restrictions through side channels and covert data access, demonstrating that permission grants do not fully constrain app behavior. This literature establishes that Android's design directly shapes user expectations and their ability to revise prior decisions, providing critical context for our investigation of silent intra-group expansion on Android~16.
	
	\noindent\textbf{Custom Permissions and Cross-App Access.}
	Custom permissions introduce developer-defined access control that operates outside AOSP's dangerous permission model. Tuncay et al.~\cite{Tuncay} showed that inconsistently defined or weakly protected custom permissions enable privilege-escalation attacks through confused deputy vulnerabilities. Li et al.~\cite{liAndroidCustomPermissions2021} systematized fuzzing of permission enforcement paths, finding fragile logic around declaration, default protection levels, and runtime checks. Enck et al.~\cite{Enck} developed TaintDroid to track sensitive data flows across app boundaries, revealing how inter-app communication can leak private information. Gamba et al.~\cite{Gamba2024Mules,gambaAnalysisPreinstalledAndroid2020} demonstrated that \perm{normal}-level custom permissions frequently guard exported components handling sensitive information, enabling cross-app data access without corresponding AOSP dangerous permissions. Their responsible disclosure to Google received acknowledgment but no remediation, with the platform citing openness concerns. These studies show that auto-granted custom permissions effectively become symbolic labels, disconnecting user expectations from actual access control outcomes. Our work extends this line by confirming these patterns persist in Android~16, dynamically validating cross-developer exploitation, and quantifying prevalence at ecosystem scale.
	
	\noindent\textbf{Wearable and IoT Companion-App Permission Studies.}
	A related line of work studies permission-mediated exposure in wearable and IoT companion-app ecosystems. Mujahid et al.~\cite{mujahid2018studying,mujahid2017detecting} analyze permission mismatches across Wear apps, Tileria et al.~\cite{tileria2020wearflow} apply taint analysis to Wear OS companion apps, Yeke et al.~\cite{yeke2024wearsmydata} examine the Android/Wear OS runtime permission model, and Neupane et al.~\cite{Neupane2022OnTD} extend this line to IoT mobile companion apps. The common focus in that literature is a same-developer companion ecosystem and information flow between phone and paired-device components. Our analysis instead targets cross-developer apps co-installed on a single Android device, where the relevant path is Android IPC via \texttt{ContentResolver.query}. Because the exploiting and exploitable apps never share an address space, there is no intra-process data flow to taint-track; the consent violation occurs at the IPC boundary the moment the unrelated app issues the query, so we verify reachability end-to-end via Probe Apps on Android~16 rather than inferring it from a companion-app flow graph. The threat model is therefore unauthorized cross-app reads enabled by within-group auto-grant and \perm{normal} custom permissions, rather than companion-app information flow across trusted endpoints.

	\section{Ethical Considerations and Responsible Disclosure}
	\label{sec:ethics}
	Our study identifies apps that expose sensitive data to unrelated developers through \perm{normal}-level custom permissions; we withheld app identities and followed coordinated disclosure. Of the 307 cross-developer pairs, 102 exploitable apps are currently listed on Google Play and were reported by email to the Google Android Security Team; apps no longer on Play or only on third-party markets were reported by email through the same channel. We also emailed the developers of the exploitable apps whose providers returned concrete data-type indicators: 43 developers acknowledged receipt of our report, and 14 have since confirmed that a new build addresses the issue, with no provider now protected by a \perm{normal} custom permission. At the time of writing, we have not received a substantive response from Google or a confirmed platform-level remediation. Dynamic testing ran on isolated researcher-owned devices with synthetic data; no sensitive user data was retained. The study involves no human subjects and no IRB review is applicable.

	\section{Conclusion}
	\label{sec:conclusion}

	Android's permission model is built on the premise that the user consents once and the platform enforces what was approved. Two long-documented mechanisms weaken that premise: permission groups silently broaden an app's capabilities after the user has approved any one member of the group, and \perm{normal}-level custom permissions grant any installed app access to a developer-defined permission with no user prompt and no surface in Settings to revoke it. Both mechanisms turn one-time consent into a standing authorisation that the user has no practical way to inspect, and on Android~16 they still expose contacts, messages, location, credentials, identity, and health records to unrelated developers, including on widely deployed and pre-installed software. This paper shows that returning informed consent to Android's update path is possible from outside the operating system: a lightweight transparency layer built entirely on public Android APIs surfaces each silent expansion to the user at update time and lets them act on it.
	
	\paragraph{Acknowledgments.} We thank the anonymous reviewers for their comments. This work was supported in part by the National Science Foundation under grant CNS-2211576 and by a Google Android Security and Privacy Research (ASPIRE) Award.
	
	\paragraph{Disclosure of Interests.} The authors have no competing interests to declare that are relevant to the content of this article.

	%
	%
	%
	
	\clearpage
	\appendix
	\section{Additional Permission-Group Expansion Flows}
	\label{app:topk_flows}
	
	\noindent Table~\ref{tab:topk_flows_appendix} reports the top permission-group expansion flows for the five groups not shown in the main-paper table.
	
	\begin{table}[h!]
		\centering
		\caption{Top-6 permission-group expansion flows for the remaining five groups.}
		\label{tab:topk_flows_appendix}
		\scriptsize
		\begin{minipage}[t]{0.48\linewidth}
			\centering
			\setlength{\tabcolsep}{3pt}
			\begin{tabular}{l l r}
				\toprule
				Group & Flow & Count \\
				\midrule
				STORAGE & Write Ext. $\rightarrow$ Read Ext. & 177,685 \\
				STORAGE & Write Ext. $\rightarrow$ Images & 105,909 \\
				STORAGE & Read Ext. $\rightarrow$ Images & 104,765 \\
				STORAGE & Write Ext. $\rightarrow$ Video & 66,054 \\
				STORAGE & Read Ext. $\rightarrow$ Video & 65,045 \\
				STORAGE & Read Ext. $\rightarrow$ Audio & 45,117 \\
				LOCATION & Fine $\rightarrow$ Coarse & 47,818 \\
				LOCATION & Fine $\rightarrow$ Background & 35,339 \\
				LOCATION & Coarse $\rightarrow$ Background & 33,451 \\
				LOCATION & Coarse $\rightarrow$ Fine & 22,451 \\
				LOCATION & Background $\rightarrow$ Coarse & 1,287 \\
				LOCATION & Background $\rightarrow$ Fine & 195 \\
				\bottomrule
			\end{tabular}
		\end{minipage}\hfill
		\begin{minipage}[t]{0.48\linewidth}
			\centering
			\setlength{\tabcolsep}{3pt}
			\begin{tabular}{l l r}
				\toprule
				Group & Flow & Count \\
				\midrule
				NEARBY DEVICES & BT Connect $\rightarrow$ BT Advert. & 2,818 \\
				NEARBY DEVICES & BT Scan $\rightarrow$ BT Advert. & 2,442 \\
				NEARBY DEVICES & BT Connect $\rightarrow$ BT Scan & 1,933 \\
				NEARBY DEVICES & BT Scan $\rightarrow$ BT Connect & 427 \\
				NEARBY DEVICES & BT Advert. $\rightarrow$ BT Connect & 59 \\
				NEARBY DEVICES & BT Advert. $\rightarrow$ BT Scan & 48 \\
				CALENDAR & Write $\rightarrow$ Read & 839 \\
				CALENDAR & Read $\rightarrow$ Write & 262 \\
				SENSORS & Body $\rightarrow$ Body BG & 49 \\
				SENSORS & Body BG $\rightarrow$ Body & 2 \\
				\bottomrule
			\end{tabular}
		\end{minipage}
	\end{table}
	
	\section{Per-Group Expansion Volume}
	\label{app:per_group}
	
	\noindent Table~\ref{tab:per_group} breaks down permission-group expansion volume by group across the 2{,}244{,}575 multi-version apps analysed in Section~\ref{subsec:group_method}. Values are independent of the VT threshold; per-group malware concentration is discussed in the body.
	
	\begin{table}[h!]
		\centering
		\scriptsize
		\setlength{\tabcolsep}{4pt}
		\caption{Permission-group expansion volume over 2{,}244{,}575 multi-version apps.}
		\label{tab:per_group}
		\begin{tabular}{lrrr}
			\toprule
			Permission Group & Expanding Apps & Expansions & \% of Total \\
			\midrule
			STORAGE          & 283{,}875 & 677{,}705 & 73.0 \\
			LOCATION         & 96{,}947  & 140{,}541 & 15.1 \\
			PHONE            & 43{,}682  & 58{,}976  & 6.4 \\
			CONTACTS         & 22{,}427  & 30{,}591  & 3.3 \\
			SMS              & 5{,}361   & 11{,}463  & 1.2 \\
			NEARBY\textunderscore DEVICES  & 4{,}671 & 7{,}775 & 0.8 \\
			CALENDAR         & 1{,}087   & 1{,}101   & 0.1 \\
			CALL\textunderscore LOG        & 234 & 249 & 0.0 \\
			SENSORS          & 51        & 51        & 0.0 \\
			\midrule
			\textbf{Total}   & \textbf{381{,}026} & \textbf{928{,}452} & \textbf{100.0} \\
			\bottomrule
		\end{tabular}
	\end{table}
	
	\clearpage
	\section{Stratified Malware Association}
	\label{app:stratified_or}
	
	\noindent Table~\ref{tab:stratified_or} reports quartile-stratified odds ratios for the malware-expansion association when controlling for each app's maximum declared permission count.
	
	\begin{table}[h!]
		\centering
		\scriptsize
		\caption{Quartile-stratified malware odds ratios (VT $\geq$ 20) for expansion status, controlling for total permission count.}
		\label{tab:stratified_or}
		\begin{tabular}{lrr}
			\toprule
			Max Permission Quartile & Apps (n) & OR \\
			\midrule
			Q1 (1--8)     & 682{,}493 & 0.93 \\
			Q2 (9--12)    & 451{,}029 & 0.78 \\
			Q3 (13--23)   & 576{,}347 & 0.70 \\
			Q4 (24+)      & 534{,}706 & 2.06 \\
			\midrule
			Pooled Mantel-Haenszel & 2{,}244{,}575 & 1.05 [1.02, 1.08] \\
			\bottomrule
		\end{tabular}
	\end{table}
	
	%
	%
	%
	
	\bibliographystyle{splncs04}
	\bibliography{cpreference}

\end{document}